\documentclass[twocolumn,showpacs,aps,prb,superscriptaddress,floatfix]{revtex4}
\usepackage{graphicx}% Include figure files
\usepackage{dcolumn}% Align table columns on decimal point
\usepackage{bm}% bold math
\usepackage{amssymb}

\begin{document}

\preprint{}

\title{A robust but disordered collapsed-volume phase in a cerium alloy under the application of pulsed magnetic fields} 

\newcommand{\celath}{Ce$_{0.8}$La$_{0.1}$Th$_{0.1}$}
\newcommand{\ceth}{Ce$_{0.9}$Th$_{0.1}$}
\newcommand{\gtoa}{$\gamma - \alpha$}

%\draft 
\author{J.P.C. Ruff}
\affiliation{Advanced Photon Source, Argonne National Laboratory, Argonne, Illinois, 60439, U.S.A.}
\author{Z. Islam} 
\affiliation{Advanced Photon Source, Argonne National Laboratory, Argonne, Illinois, 60439, U.S.A.}
\author{R.K. Das} 
\affiliation{Advanced Photon Source, Argonne National Laboratory, Argonne, Illinois, 60439, U.S.A.}
\author{H. Nojiri} 
\affiliation{Institute for Materials Research, Tohoku University, Katahira, Sendai 980-8577, Japan}
\author{J.C. Cooley} 
\affiliation{Los Alamos National Laboratory, Los Alamos, New Mexico, 87545, U.S.A.}
\author{C.H. Mielke} 
\affiliation{Los Alamos National Laboratory, Los Alamos, New Mexico, 87545, U.S.A.}

\begin{abstract} % insert abstract here 

We report synchrotron x-ray powder diffraction measurements of {\celath} subject to pulsed magnetic fields as high as 28 Tesla.  This alloy is known to exhibit a continuous volume collapse on cooling at ambient pressure, which is a modification of the {\gtoa} transition in elemental cerium.  Recently, it has been suggested on the basis of field-cooled resistivity and pulsed field magnetization measurements that the volume collapse in this alloy can be suppressed by the application of magnetic fields.  Conversely, our direct diffraction measurements show a robust collapsed phase, which persists in magnetic fields as high as 28 Tesla.  We also observe nanoscale disorder in the collapsed phase, which increasingly contaminates the high temperature phase on thermal cycling.

\end{abstract} 

\maketitle 
% \narrowtext
% \twocolumn
% body of paper here

\section{Introduction}

Elemental cerium is stark example of the complicating effects of magnetism in metals.  In principle a simple system, cerium at room temperature and ambient pressure presents a monatomic Bravais lattice (the face-centered-cubic $\gamma$ phase).  However, under the application of modest pressures, cerium undergoes a spectacular first-order transition to a low temperature $\alpha$ phase.  Both $\alpha$ and $\gamma$ are isostructural f.c.c. phases, but differ by $\sim 15 \%$  in volume, and have markedly different magnetic susceptibilities and resistivities.\cite{bestreview}  There are competing explanations for the transition, which variously invoke a ``Kondo Volume Collapse''\cite{KVC} mechanism, a ``Mott Transition'',\cite{MT} or some entropic mechanism\cite{entropy,jpcm_celath}.  In all cases however, it is generally agreed that the large volume $\gamma$ phase contains relatively larger localized magnetic moments (evinced by the higher susceptibility) and relatively less itinerant electrons (evinced by the higher resistivity).  Beyond that, the nature of the transisiton remains contentious.

One of the main complications in the study of cerium is that although the {\gtoa} transition occurs on cooling over a wide range of pressures, it is avoided at ambient pressure.  In that case, an intervening dhcp phase forms from the $\gamma$ phase, and transforms into the $\alpha$ phase only at lower temperature.\cite{bestreview}  There are commonly two methods used to resolve this problem.  Measurements are either performed under pressure on pure cerium, or else the cerium is alloyed with other elements known to suppress the dhcp phase (typically thorium).  Results from these two avenues of investigation are not always consistent - for example, it has been claimed that the phonon entropy plays little role in the {\gtoa} transition in a {\ceth} alloy,\cite{neutron1,neutron2} while studies on pure cerium under pressure have returned the opposite result.\cite{puretron,PNASxray}  Nevertheless, the alloying of cerium has generated a family of related compounds, in which the order of the transition, the magnitude of the volume collapse, and the transition temperature can be smoothly varied.\cite{fiskthompson,lashtricrit}

One such alloy, {\celath}, is of particular interest in regards to spin-lattice coupling at the phase transition.  Alloying with thorium suppresses the dhcp phase, while alloying with lanthanum causes the transition to become continuous and supresses the critical temperature.\cite{fiskthompson, lashtricrit}  It has been reported that cooling {\celath} in high magnetic fields can suppress the resistivity change associated with the volume collapse to lower temperature, and that the application of pulsed magnetic fields can induce a large moment below the critical temperature.\cite{jpcm_celath}  These measurements map out a phase boundary for the {\gtoa} transition as a function of magnetic field, and suggest that the volume collapse can be fully suppressed by $\mu_{0} H \lesssim 56$ Tesla.  Here we report a direct study of the structure of {\celath}, using synchrotron x-ray powder diffraction in pulsed magnetic fields as high as 28 Tesla.  Lattice parameter and structural correlation length are extracted throughout the $H - T$ phase diagram.

\section{Experiment}

Samples of {\celath} were prepared at Los Alamos by arc-melting elemental cerium, lanthanum and thorium in a zirconium gettered argon atmosphere.  The button was melted and flipped 15 times to ensure mixing.  The final melt was cooled in a round-bottomed trough, which was 5 mm wide.  The resulting $\sim$ 25 mm long sample was unidirectionally rolled $\sim$ 10 times to achieve a finished thickness of 50 microns.  A section of the as-rolled material was heat treated at 450$^\circ$C for 1 hour to remove the rolling texture before a final sample, 2 mm in diameter, was cut from the sheet with a razor blade.  The sample was subsequently mounted inside a blind hole in a sapphire bracket, and sealed with a sapphire cap and stycast epoxy, to ensure good thermal contact and prevent oxidization.  The sapphire bracket was anchored to the cold finger of the double-funnel solenoid pulsed magnet at the Advanced Photon Source (APS), which facilitates diffraction at up to 22$^{\circ}$ of scattering angle, to temperatures as low as $\sim 5$ K and in magnetic fields as high as $\sim$ 30 Tesla.\cite{solenoid1,solenoid2}  This system features a resistive magnet coil (of Tohoku design) in a liquid nitrogen bath, connected to a 40 kJ capacitor bank.  The bank is capable of discharging a 3 kV charge through the coil in a $\sim 6$ ms half-sine wave pulse, generating a peak current of $\sim 10$ kA and a peak magnetic field of $\sim 30$ Tesla.  The heat generated in the resistive coil during a pulse must be allowed to dissipate before the system can be pulsed again.  At peak field values, this requires a wait time $\gtrsim$ 8 minutes between pulses.  The sample to be studied is both thermally and vibrationally isolated from the coil.  This new instrument offers a complementary measurement geometry to the APS split-pair pulsed magnet developed previously.\cite{firstrsi,ttopmag}  

\begin{figure} 
\centering 
\includegraphics[width=8.1cm]{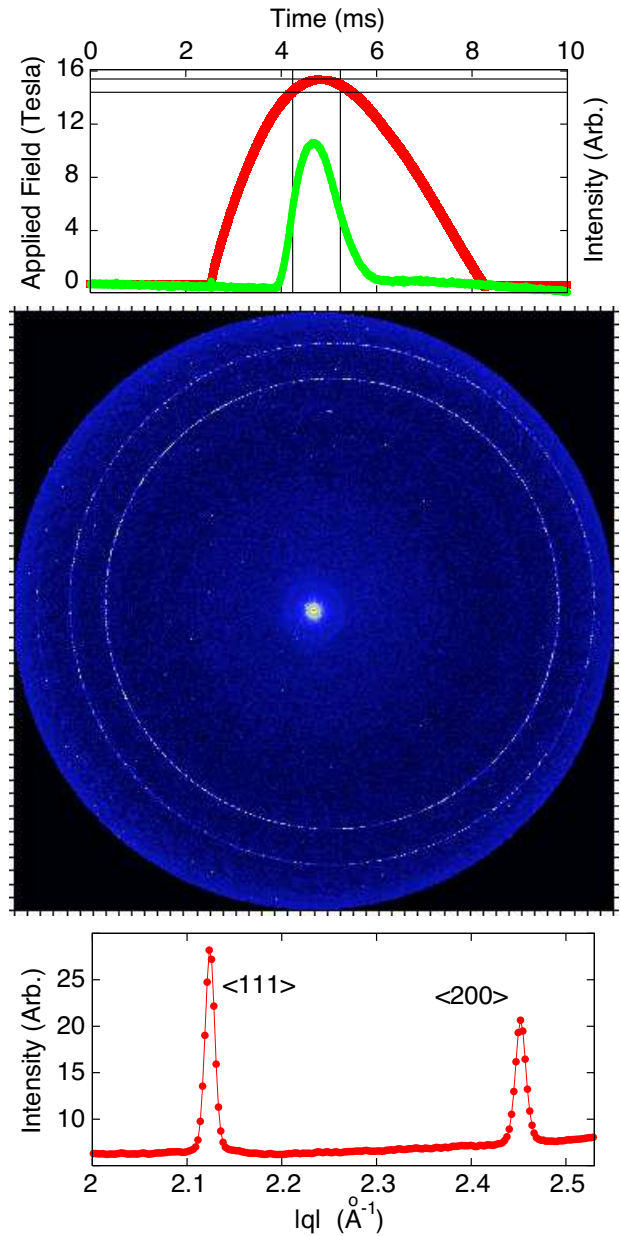}
\caption {Top:  The inicident x-ray pulse (the narrow pulse) is timed to arrive during the magnetic field peak (broader pulse).  In this case, a 5.75 ms magnetic field pulse to ~15.4 Tesla weighted by an x-ray exposure with a 1 ms fullwidth gives an average magnetic field value of $\sim$ 14.9 Tesla, integrated over  $\pm$ 0.5 Tesla for the exposure.  Middle:  Raw data collected on the image plate for a single 1 ms exposure.  Bottom:  The same data, radially integrated and converted to $|\bf{q}|$.}
\label{fig:1}
\end{figure}

Experiments were performed at the 6ID-B station at the APS.  30 keV photons were selected using a Si(111) monochromator, with the mirror removed to provide an unfocused beam.  The monochromator was detuned so as to reduce higher harmonic contamination.  The incident beam size was defined by slits, with an illuminated area on sample of 0.2 mm X 1.0 mm.  A $\sim 1$ ms pulse of x-rays was defined using a pair of fast platinum shutters, and the arrival of this pulse at the sample was synchronized with the peak of the magnetic field pulse.  Scattered x-rays were recorded on a two-dimensional image plate detector.  In this configuration, our system is similar to the pulsed magnet instrument in operation at the European Synchrotron Radiation Facility (ESRF).\cite{detlefs_ins}  To calibrate our instrument, we repeated the measurements\cite{solenoid2} of Detlefs \textit{et al.} on polycrystalline TbVO$_4$ in pulsed magnetic fields, as previously measured by the ESRF instrument.\cite{detlefs_tbvo}  We reproduced the reported high field peak splitting in the tetragonal phase of TbVO$_4$,\cite{detlefs_tbvo} validating our measurement scheme.\cite{solenoid1,solenoid2}  We performed our measurements of {\celath} with the axis of the solenoid coincident with the incident beam, and the detector positioned so as to capture the majority of the solenoid exit window.  In this way, at 30 keV the first two powder rings (corresponding to the $<$111$>$ and $<$200$>$ peaks) are completely captured by the detector, allowing a full radial integration and excellent counting statistics even for a single 1 ms exposure.  The current through the solenoid and the x-ray flux incident on the sample were monitored as a function of time by tracing the output of a high-resolution current monitor and an air-filled ion chamber on a digital storage oscilloscope.  A typical trace is shown in Fig. 1, along with data from a single exposure on the image plate.  In order to extract the correlation length and lattice parameter from these diffraction patterns, each of the two lowest lying peaks was fit individually, generating two independent sets of data, which were subsequently compared.  In all cases, the properties extracted from the two peaks aggreed within the errorbars of the measurement, and the quantities quoted in the following are the average of the two.

\section{Zero-Field Results}

Powder diffraction patterns at room temperature show a single phase material, with symmetry and lattice constant consistent with the $\gamma$ phase of {\celath}.  On cooling, we observe the first two Bragg peaks to shift to higher $|\bf{q}|$, while also broadening by a factor of $\sim 2$ and exhibiting a reduction in peak intensity (see Fig. 2).  This is consistent with a collapse in volume, accompanied by a striking reduction in the structural correlation length.   It is interesting to note that we observe a single phase at all temperatures, in contrast to recent neutron diffraction measurements on {\ceth} which reported a coexistence of both the $\alpha$ and $\gamma$ phases over a wide range of temperatures.\cite{neutron2}  Our sample was initially cooled directly to 8K to characterize the magnitude of the volume collapse, then warmed to 250 K while collecting data.  The volume was observed to approach the $\gamma$ phase value quite gradually.  Next we cooled slowly and measured the lattice parameters with a fine point density.  The resulting continuous but hysteretic volume collapse curves are plotted in Fig. 3.  Surprisingly, on our second cooling we found that the absolute size of the volume collpase was diminished - thermal cycling of the sample is detrimental to the phase transition.

\begin{figure} 
\centering 
\includegraphics[width=8.5cm]{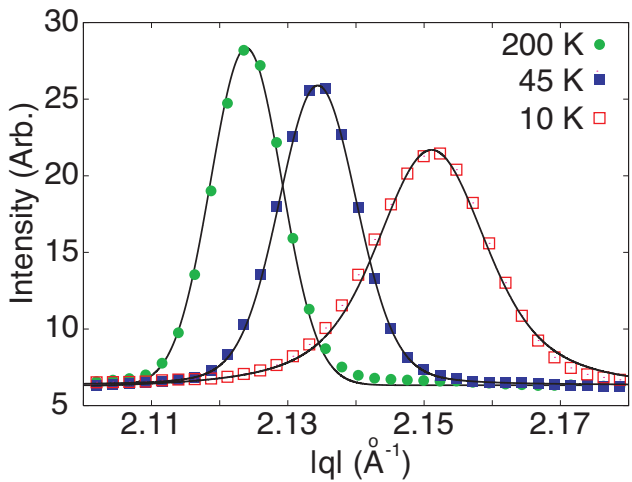}
\caption {Zero-field temperature dependence of the $<$111$>$ Bragg reflection.  The peak shifts to higher $|\bf{q}|$ and broadens on cooling.  The sample is a single phase at all temperatures.  Resolution limited peaks at high temperature are well described by Gaussians, while a Voigt shape at low temperature is consistent with an Ornstein-Zernike scattering function with a short correlation length (solid lines).}
\label{fig:2}
\end{figure}

\begin{figure} 
\centering 
\includegraphics[width=8.5cm]{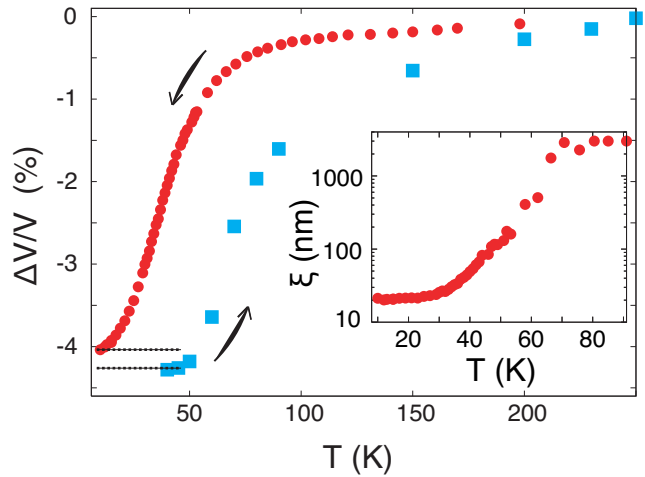}
\caption {Zero-field volume collapse in {\celath}.  Sample was warmed from 8 K to 250 K ($\blacksquare$), then cooled back to 10 K ($\bullet$).  This thermal cycling lead to a reduction in the magnitude of the volume collapse, shown by the dashed lines.  The inset shows the Ornstein-Zernike correlation length extracted from the cooling measurements (see text).  The correlated regions of the collapsed phase have average size of only $\sim 20$ nm.}
\label{fig:3}
\end{figure}

The observed peak broadening can be modelled phenomenologically, by assuming that the high temperature patterns show long-range ordered and resolution limited scattering, while the low temperature broadening arises from a reduction in the Ornstein-Zernike correlation length.  Our high temperature peaks are well described by Gaussians, and we are assuming a Lorentzian form for the scattering function (see Ref.[\onlinecite{strucfluc}] for details).  Therefore, the measured scattering at low temperature should be well described by the convolution of a Lorentzian and a Gaussian - a Voigt function.  Since the width of this Voigt function is related to the width of the Gaussian (the resolution) and the width of the Lorentzian (the inverse correlation length), the numerically extracted fullwidth of the Bragg peaks can be converted to a resolution-deconvoluted correlation length.  The solid lines in Fig. 2 demontrate the degree of agreement between these lineshapes and the measured data, and the extracted correlation length is plotted in the inset of Fig. 3.  Our resolution limit is found to be on the order of a few microns, however, the low temperature collapsed phase exhibits correlation lengths which are roughly two orders of magnitude shorter than this.  Below $\sim$ 30 K, the $\alpha$ phase correlations are seen to extend over an average range of $\sim$ 20 nm.  Since our alloy is $\sim$ 80$\%$ cerium, the mean distance between dopant atoms (either lanthanum or thorium) is on the order of 2 nm.  Thus, while the inclusion of lanthanum and thorium in the alloy does not completely hinder the formation of the $\alpha$ phase, and each correlated region on average contains multiple dopant atoms, nevertheless these atoms act to strongly disorder the sample at the nanoscale within the $\alpha$ phase.  Conversely, within the $\gamma$ phase, individual grains are well correlated over micron-sized regions despite the same level of chemical disorder.

\section{results in pulsed magnetic field}

Motivated by the results of Ref. [\onlinecite{jpcm_celath}], we next sought to characterize the effect of applied magnetic field on the volume collapse transition in {\celath}.  Currently, the available capacitance at the APS limits our measurements to applied magnetic fields $\lesssim$ 30 Tesla, well below the 56 Tesla estimated to fully suppress the {\gtoa} transition to zero Kelvin.\cite{jpcm_celath}  However, inspection of the resistivity measurements of Ref. [\onlinecite{jpcm_celath}] reveals that our highest achievable magnetic fields should be sufficient to suppress the temperature of the volume collapse inflection point by $\sim$ 8 K.  We can maximize our sensitivity to this effect by cooling to the inflection point of the volume collapse ($\sim$ 35 K) and pulsing to high magnetic field.  The slope of $\frac{\Delta V}{V}$ is large at this temperature, and therefore an 8 K shift in the inflection point should cause a $\sim$ 0.7 $\%$ change in volume at 35 K, well within our resolution.  However, as can be seen from Fig. 4, we failed to observe this effect.  After cooling to 35 K, we collected five diffraction patterns at zero field, which served to define the zero field Bragg peak positions to an accuracy better than $\sim 10^{-3} \AA^{-1}$.  Next, we collected ten patterns in pulsed field, with the average applied field over the x-ray exposure being $\sim 28 \pm 1$ Tesla.  We observed no measurable change in the Bragg peak positions in any individual pattern.  Nor did we observe any cumulative effect from multiple high-field pulses.  Our measurements therefore constrain any volume changes at (35 K, 28 T) to be less than $\sim$ 0.07 $\%$, a full order of magnitude smaller than what would be expected.\cite{jpcm_celath} 

\begin{figure} 
\centering 
\includegraphics[width=8.5cm]{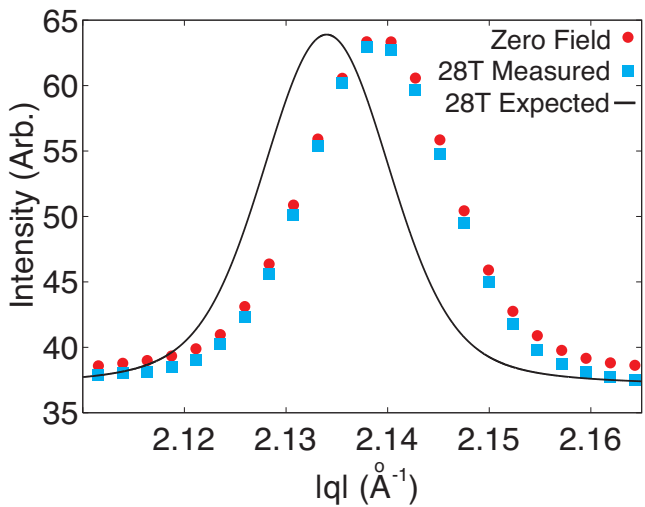}
\caption {Observed and expected effects of 28 Tesla pulsed magnetic fields on the $<$111$>$ Bragg peak at T=35 K.  There is no discernable change in the position or width of the peak.  The expected curve is calculated based on the position and width change that would arise from an 8 K shift in the transition temperature, consistent with the results of Ref. [\onlinecite{jpcm_celath}].}
\label{fig:4}
\end{figure}

\begin{figure} 
\centering 
\includegraphics[width=8.5cm]{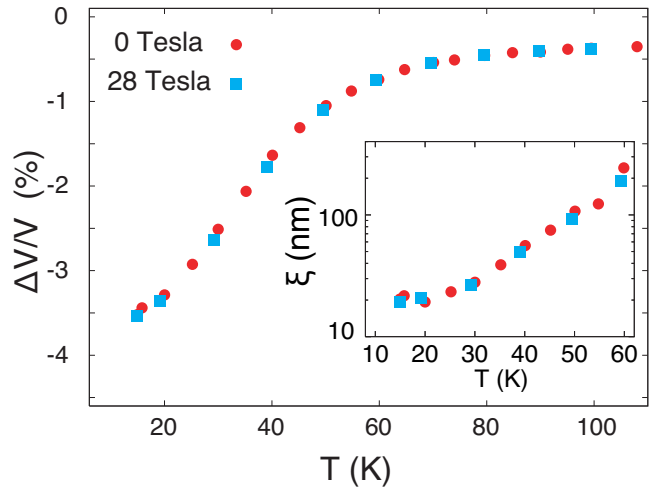}
\caption {Volume collapse ($\Delta V/V$) and Ornstein-Zernike correlation length ($\xi$), comparing zero field cooled measurements and measurements where the system is repeatedly pulsed to 28 T on cooling.  No field effect is seen.}
\label{fig:5}
\end{figure}

It is important to remember that the evidence for field induced suppression of the volume collapse in {\celath} for magnetic fields below 30 Tesla comes from resistivity curves, measured by cooling in constant applied magnetic fields.  This is procedurally distinct from the zero-field-cooled pulsed measurements reported here.  Discrepancy between field-cooled and zero-field-cooled behaviour is a hallmark of spin glass materials, wherein chemical disorder and competing interactions collude to freeze in disordered ground states.\cite{spinglassrev}  We have already shown that chemical disorder in our alloy has a strong effect on the development of $\alpha$-phase correlations (see Fig. 3).  It is therefore tempting to sugest that there may be some glassy behavior at play in {\celath}.  Direct comparison between field cooled and zero field cooled measurements are therefore highly desirable, however, a truly field-cooled measurement is not achievable with pulsed magnets.  We have attempted to approximate one by slow cooling, and pulsing in regular intervals.  In order to minimize the obscuring effects of thermal cycling, we performed two identical cooling curves, one after the other, warming to 280 K once to reset the system.  For each of these measurements, we cooled at a controlled rate of 1 K$/$min from 280 K.  During the first of these cooldowns, we pulsed to 28 T at 10 min (10 K) intervals.  During the second cooldown, we collected zero-field data in the same manner, albeit more frequently since there was no need to wait for the coil to cool down. The results are plotted in Fig. 5.  Clearly, magnetic fields applied in this way are no more effective at suppressing the volume collapse than the zero-field-cooled methods shown in Fig. 4. 

We are therefore forced to conclude that the available magnetic fields were insufficient to induce any structural effects in {\celath}.  If the {\gtoa} transition were continuous for this alloy (as claimed in Ref. [\onlinecite{fiskthompson}]), and the magnetic field induced suppression occured as claimed in Ref. [\onlinecite{jpcm_celath}], then we would have expected to see a clear shift in $|\bf{q}|$ as denoted in Fig. 4 for the zero-field-cooled measurements, and a clear suppression in temperature should have occurred for the repetitively pulsed measurements reported in Fig. 5.  Conversely, if the {\gtoa} transition were first order for this alloy (as claimed in Ref. [\onlinecite{lashtricrit}]), then our measurements imply that the phase boundary was never crossed in our cooling curve.  This would require a steeper increase of the critical field on lowering temperature than was reported in Ref. [\onlinecite{jpcm_celath}.]  It is also possible that the alloy is dynamically inhibited on the timescale of the magnetic field pulse, which would be consistent with a glassy behavior.  Finally, it is possible that the effects reported in Ref. [\onlinecite{jpcm_celath}] were not intrinsic to {\celath}.  Clearly, the issue of field induced suppression of the {\gtoa} transition requires further investigation in elemental cerium and its alloys.

\section{effect of thermal cycling}

As has been noted in the preceeding sections, we have observed a detrimental effect due to thermal cycling.  Our initial measurements on as-grown samples revealed well-correlated grains on the order of microns in size.  On first cooling, the volume collapsed $\alpha$ phase was seen to exhibit nanoscale disorder, and on second cooling the absolute size of the volume collapse had diminished (see Fig. 3 and discussion in Section III).  Over the course of our week-long experiment, we observed the $\alpha$-phase volume to gradually increase with each thermal cycling, while the $\gamma$-phase volume was seen to gradually decrease.  In addition, while the $\alpha$-phase consistently showed nanoscale disorder with correlation lengths of $\sim$ 20 nm, the $\gamma$ phase correlations were observed to shrink, from micron-sized grains in the as-grown sample to correlation lengths of only $\sim$ 150 nm after 11 thermal cycles.  Zero-field cooling curves from the second, seventh, and eleventh cycle are plotted in Fig. 6, to illustrate the effect.  It is likely that the strains associated with the large volume change and spatial variations in doping,\cite{fiskthompson} which are manifest in the $\alpha$-phase disorder, act to introduce new domain boundaries at low temperature, which thereafter persist as defects.  This effect is insidious, since the suppression by thermal cycling we observe here could easily be mistaken for suppression by an external perturbation, if that perturbation was varied monotonically during the course of an experiment.  Hypothetically, had we measured cooling curves through the volume collapse in DC magnetic fields which were monotomically increased during the course of our measurements, we may easily have misinterpreted the gradual buildup of thermal cycling damage as a magnetic field induced effect.  Therefore, it is important to take extra care when mapping out phase diagrams in cerium alloys.  More generally, these measurements highlight the need for careful consideration of strain driven effects in all materials studies, which can often obfuscate results.

\begin{figure}
\centering 
\includegraphics[width=8.5cm]{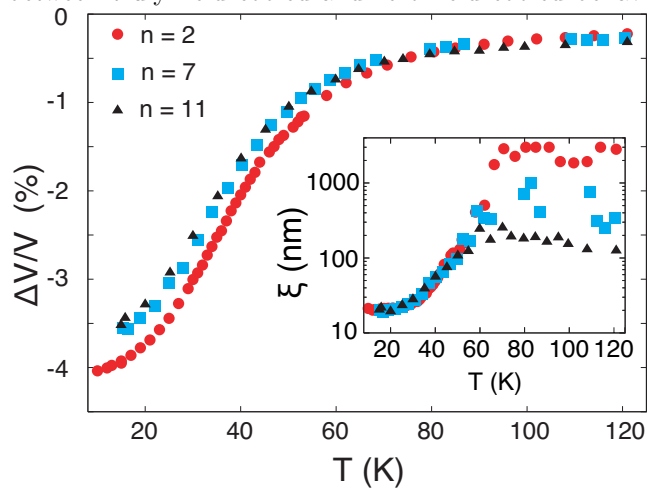}
\caption {Gradual contamination by nanoscale disorder, induced by thermal cycling through the {\gtoa} transition.  The effects are clear in both the magnitude of the volume collapse ($\Delta V/V$) and the Ornstein-Zernike correlation length ($\xi$).  All data was collected on cooling in zero magnetic field.  The three data sets represent the second ($\bullet$), seventh ($\blacksquare$), and eleventh ($\blacktriangle$) thermal cycle.}
\label{fig:6}
\end{figure}

\section{conclusions} 

In this article, we have reported a synchrotron x-ray diffraction study of the volume collapse transition in {\celath}, as a function of temperature and pulsed magnetic field.  In this alloy, the {\gtoa} transition is smooth but hysteretic, characterized by a $\sim 4 \%$ volume change and a dramatic reduction in structural correlation length.  The collapsed $\alpha$ phase is shown to be unaffected by pulsed magnetic fields $\lesssim$ 28 Tesla, both in zero-field cooled conditions and under repetitive pulsing on cooling.  It is suggested that a deviation between truly field cooled and zero field cooled behavior in {\celath} may account for the disagreement between our measurements and previous bulk studies.\cite{jpcm_celath}  This implies a glassy character to the low temperature phase.  We have also shown that repeated thermal cycling through the {\gtoa} transition acts to disorder the sample on nanometer length scales, while suppressing the magnitude of the volume collapse transition.  These measurements may constitute a microscopic measure of what has been generally refered to as ``sluggish dynamics'' in cerium alloys.\cite{jpcm_celath,fiskthompson}

It is our hope that these results will advance understanding of the exotic {\gtoa} transition in elemental cerium and its alloys.  Specifically, we believe the data presented here highlights the strong effects of disorder in these systems.  We gratefully acknowledge fruitful discussions with J. Lashley and B. Toby.  Use of the Advanced Photon Source is supported by the DOE, Office of Science, under Contract No. DE-AC02-06CH11357.  Pulsed magnet collaborations between Argonne and Tohoku University are supported by the ICC-IMR.   HN acknowledges KAKENHI No. 23224009 from MEXT.  JPCR aknowledges the support of NSERC of Canada.

% now the references. delete or change fake bibitem. delete next three
%   lines and directly read in your .bbl file if you use bibtex.

% that you will use with ref{} command in the braces of the \label{} command.
%

% Uncomment the following to put the figures into the latex file.

%\begin{figure}
%\centering
%\includegraphics[width=8.5cm]{fig4_nolines_tilt.ps}
%\caption{Enter caption here.}
%\label{fig4}
%\end{figure}

%\begin{figure}
%\centering
%\includegraphics[angle=0,origin=c,width=8cm]{all4_2.ps}
%\caption{Caption}  
%\label{fig2}
%\end{figure}

%\begin{figure}
%\centering
%\includegraphics[angle=0,origin=c,width=8cm]{th2thputtogether2.ps}
%\caption{Caption}
%\label{fig3}
%\end{figure}

%\begin{figure}
%\centering
%\includegraphics[angle=0,origin=c,width=8cm]{dspace.ps}
%\caption{Caption}
%\label{fig4}
%\end{figure}

\end{document}